\documentclass[preprint2]{aastex}

\def\kms{\mbox{km/s}}
\def\kpc{\mbox{kpc}}
\def\kpch{\mbox{$h^{-1}$kpc}}
\def\hubble{\mbox{km sec$^{-1}$ Mpc$^{-1}$}}

\def\mpc{\mbox{Mpc}}
\def\Mpc{\mbox{Mpc}}
\def\mpch{\mbox{$h^{-1}$Mpc}}
\def\Mpch{\mbox{$h^{-1}$Mpc}}
\def\msun{\mbox{M$_\odot$}}
\def\msunh{\mbox{$h^{-1}$M$_\odot$}}

\def\sige{\mbox{$\sigma_8$}}
\def\sig3dm{\mbox{$\sigma_{3D,dm}$}}
\def\etal{et al.}
\def\pvb{b_{v,12}}
\def\Vmax{\mbox{$V_{\rm max}$}}

\def\mathnew{\mathsurround=0pt}
\def\ref{\par\noindent\hangindent=2pc \hangafter=1 }
\def\simov#1#2{\lower .5pt\vbox{\baselineskip0pt
    \lineskip-.5pt\ialign{$\mathnew#1\hfil##\hfil$\crcr#2\crcr\sim\crcr}}}  
\def\simgreat{\mathrel{\mathpalette\simov >}}
\def\simless{\mathrel{\mathpalette\simov <}}
\def\'#1{\ifx#1i{\accent"13\i}\else{\accent"13#1}\fi}
\def\eg{e.g.,}

\shorttitle{Velocity Bias}
\shortauthors{Col\'in et al.}
\begin{document}

\title{Velocity bias in a $\Lambda$CDM model}

\author{Pedro Col\'in}
\affil{Instituto de Astronom\'ia, Universidad Nacional Aut\'onoma
de M\'exico, C.P. 04510, M\'exico, D.F., M\'exico}

\author{Anatoly A. Klypin, and Andrey V. Kravtsov}
\affil{Astronomy Department, New Mexico State University, Box 30001, Department
4500, Las Cruces, NM 88003-0001}

\begin{abstract}

We use a high resolution $N$-body simulation to study the velocity bias of
dark matter halos, the difference in the velocity fields of dark
matter and halos, in a flat low-density $\Lambda$CDM model.  The high
force, $2\kpch$, and mass, $10^9h^{-1} {\rm M_{\odot}}$, resolution
allows dark matter halos to survive in very dense environments of
groups and clusters making it possible to use halos as galaxy
tracers.  We find that the velocity bias $\pvb$ measured as a ratio of
pairwise velocities of the halos to that of the dark matter evolves
with time and depends on scale. At high redshifts ($z \sim 5$) halos
move generally faster than the dark matter almost on all scales:
$\pvb(r)\approx 1.2, r>0.5\Mpch$. At later moments the bias decreases
and gets below unity on scales less than $r\approx 5\Mpch$:
$\pvb(r)\approx (0.6-0.8)$ at $z=0$.  We find that the evolution of
the pairwise velocity bias follows and probably is defined by the
spatial antibias of the dark matter halos at small scales.  One-point
velocity bias $b_v$, defined as the ratio of the rms velocities of halos
and dark matter, provides a more direct measure of the difference in
velocities because it is less sensitive to the spatial bias.  We
analyze $b_v$ in clusters of galaxies and find that halos are ``hotter''
than the dark matter: $b_v=(1.2-1.3)$ for $r=(0.2-0.8)r_{vir}$,
where $r_{vir}$ is the virial radius.  At larger radii, $b_v$
decreases and approaches unity at $r=(1-2)r_{vir}$. 
We argue that dynamical
friction may be responsible for this small positive velocity bias
($b_v > 1$) found in the central parts of clusters. 
We do not find significant systematic difference in
the velocity anisotropy of halos and the dark matter. The dark matter the velocity anisotropy can be approximated
as $\beta(x) =0.15 +2x/(x^2+4)$, where distance $x$ is measured in
units of the virial radius.

\end{abstract}  


\section{Introduction}

Peculiar velocities of galaxies arise due to the gravitational pull of
surrounding overdense regions and therefore reflect the underlying
density field.  The statistical study of galaxy velocities is
important in cosmology since it can be used as a tool to constrain
cosmological models. The connection between theoretical predictions
and the observed statistics usually requires an additional quantity:
the difference between galaxy and dark matter velocities, termed the
velocity bias.  The situation with predictions of the velocity bias is
rather confusing. There is a wide range of estimates of the velocity
bias. Values change from strong antibias with galaxies moving twice
slower than the dark matter \citep{GB94, KHPR}, to almost no
bias (Klypin et al. 1998 (KGKK); Ghigna et al. 1998), to slight positive bias
\citep{Diaferio98, OH99}. 
Following Carlberg (1994) and Summers, Davis, \& Evrard (1995) we
distinguish two forms of the velocity bias. The one-point velocity bias
$b_v$ is defined as the ratio of the rms velocity of galaxies or
galactic tracers to that of the dark matter:
\begin{equation}
b_v = \frac{\sigma_{\rm gal}}{\sigma_{\rm DM}},
\end{equation}
\noindent where the rms velocity $\sigma$ is estimated on some
scale. Traditionally, this measure of the velocity bias is used for
clusters of galaxies. Two-particle or pairwise velocity bias
$b_{v,12}$ compares the relative velocity dispersion in pairs of objects
separated by distance $r$:
\begin{equation}
b_{v,12} = \frac{\sigma_{\rm g, g}(r)}{\sigma_{\rm dm, dm}(r)}.
\end{equation}
\noindent
The pairwise velocity dispersion (PVD) was often used to complement the
analysis of the two-point spatial correlation function. At small
scales, the cosmic virial theorem \citet{Peebles80} predicts that the
PVD of galaxies should be proportional to the product of the mean
density of the universe and the two-point correlation function.  The
PVD of galaxies has been estimated for the CfA catalog (Davis \&
Peebles 1983; Zurek \etal 1994; Somerville, Davis \& Primack 1997) and
recently for the Las Campanas Redshift Survey by Landy, Szalay, \&
Broadhurst (1998) and Jing, Mo \& B\"{o}rner (1998). The latter two
studies gave $363 \pm 44 \kms$ and $570
\pm 80 \kms$, respectively, for a $1\Mpch$ separation.  Jing \&
B\"{o}rner (1998) show
that the discrepancy between these two studies is due to the difference
in treatment of the infall velocities. The value of $\sigma_{\rm g, g}$
as well as the infall velocities depend on which regions (clusters or
field) are included in the surveyed sample.

The PVD of the dark matter, $\sigma_{\rm dm,dm}$, has also been
estimated for a variety of cosmological models (\eg\ Davis et al. 1985; 
Carlberg \& Couchman 1989; Carlberg, Couchman \& Thomas 1990; Klypin
\etal 1993; Col\'in, Carlberg, \& Couchman 1997; Jenkins et
al. 1998). If galaxies were a random sample of the mass distribution,
we would expect that $\sigma_{\rm g,g}$ were approximately equal to
$\sigma_{\rm dm,dm}$. Davis et al. (1985) showed that in this case an
$\Omega_0 = 1$ model with $\sige = 1$ produces a PVD that is too large
compared to observations. Here \sige\ is the rms of mass fluctuation
estimated with the top-hat window of radius 8\mpch. This is an example
of a model which needs some kind of bias to be compatible with the
observations.

The notion of the pairwise velocity bias $\pvb$ was introduced by
Carlberg \& Couchman (1989). They found that the dark matter had a PVD
a factor of two higher than that of the simulated ``galaxies''.  In
a further analysis, Carlberg, Couchman, \& Thomas (1990) suggested that
an $\Omega_0 = 1$ model with $\sige = 1$ could be made consistent with 
the available data for $b_{v,12} \sim 0.5$ (velocity antibias) and 
almost no spatial bias.
Estimates of the pairwise velocity bias are in the range of 0.5--0.8
\citep[][b]{CC92, CO92, GB94, ESD94, CCC97, Kauffmann98a}.
Differences between the estimates (especially the early ones) can be
attributed to some extent to numerical effects (``overmerging
problem'') and to different methods of identifying galaxy tracers. Only
recently $N$-body simulations achieved a high dynamic range in a
relatively large region of the universe necessary for a large number of
galaxy-size halos to survive in clusters and groups 
\citep[\eg KGKK,][]{ghigna98, Colin98}. 
The estimates of the pairwise velocity bias start showing a tendency
for convergence. For example, results of Kauffmann et al. (1998a, 1998b)
for a low-density model with a cosmological constant and results
presented in this paper for the same cosmological model agree
reasonably well in spite of the fact that we use very different
methods. Results point systematically to a  antibias $\pvb=
0.6-0.7$.

One-point velocity bias for clusters and groups of galaxies tells a
different story. 
Values of $b_v$ are typically larger than those for $\pvb$ and range from
0.7 to 1.1 \citep{CD91, KW93, Carlberg94, ghigna98, Frenk96, ME97, 
OH99, Diaferio98}.  Carlberg \& 
Dubinski (1991) suggested that if the pairwise velocity antibias is
significant, galaxies in clusters should have orbital velocities lower
than the dark matter. However, this may not necessarily be true. In
this paper (see also, for example, Kauffmann et al. 1998a) we argue
that galaxy tracers do not need to move slower in clusters to have the
pairwise velocity bias $\pvb < 1$. In particular, we find that while
$\pvb<1$ for halos in our study, the halos in many clusters actually
move somewhat {\em faster\/} than dark matter. Ghigna et al. (1998)
also do not detect a significant difference between the orbits of DM
particles and halos. They find that the cluster radial velocity
dispersion of halos is within a few percent of that of the DM
particles. Okamoto \& Habe (1999) used hundreds of galaxy-size halos in
their simulated cluster. They are able to compute the halo velocity
dispersion profile. Their results suggest that in the range $0.3~\Mpc
\simless r \simless 0.6~\Mpc$ halos have a velocity dispersion slightly larger
than that of the DM particles. Diaferio et al. (1998) using a
technique that combines $N$-body simulations and semi-analytic
hierarchical galaxy formation modeling also find that galaxies in
clusters have higher orbital velocities than the underlying dark
matter field. They suggest that this effect is due to the infall
velocities of blue galaxies.  We find in this paper a similar effect:
galaxy-size halos are ``hotter'' than the dark matter in clusters.

  The paper is organized as follows. In \S~2 brief
descriptions of the model, simulation, and group-finding algorithm are
given. In \S~3 the DM and halo PVDs as well as the corresponding
velocity bias are computed at four epochs. We take a sample of the most
massive clusters in our simulation and compute an average halo and DM
velocity dispersion profile. A cluster velocity bias is then defined
and computed. A discussion of the main results are presented in \S~4.
The conclusions are given in \S~5.


\section{Model, simulation, halo-finding algorithm}

We use a flat low-density model ($\Lambda$CDM) with $\Omega_0 = 1 -
\Omega_\Lambda = 0.3$ and $\sigma_8=1$. Cluster mass
estimates \citep[\eg][]{Carlberg96}, evolution of the abundance of galaxy clusters 
\citep[\eg][]{Eke98}, baryon fraction in clusters 
\citep[\eg][]{Evrard97}, and the galaxy tracer 
two-point correlation function\citep[\eg][]{Colin98, Benson99}
favor a low-density universe with $\Omega_0 \sim 0.3$ 
\citep[see also][]{RH99}.  On the other hand, various observational determinations
of $h$ (the Hubble constant in units of $100~\hubble$) converge to
values between 0.6--0.7. Our model was set to $h = 0.7$ which gives an
age for the universe of 13.4 Gyr in close agreement with the oldest
globular cluster age determinations \citep{Chaboyer98}. The
approximation for the power spectrum is that given by \citep{KH97}. 
The adopted normalization of the power spectrum is consistent with 
both the COBE observations and observed abundance of galaxy clusters. 

The Adaptive Refinement Tree code (ART; Kravtsov, Klypin \& Khokhlov
1997) was used to run the simulation, as described by Col\'{\i}n et
al. (1999). The simulation followed the evolution of $256^3$ dark matter
particles in a 60\mpch\ box which gives particle mass of $1.1 \times
10^9 \msunh$.  The peak force resolution reached in the simulation is
$\sim 2 \kpch$. The mass resolution is sufficient for resolving and
identifying galaxy-size halos with at least 30 particles.  The force
resolution allows halos to survive within regions of very high density
(as those found in groups and clusters of galaxies). In dense
environment of clusters the mass of halos is not well
defined. Therefore, we use the maximum
circular velocity
\begin{equation} 
	V_{\rm max} =\Biggl(\frac{GM(<r)}{r}\Biggr)^{1/2}_{\rm max},
\end{equation}
\noindent where $M(<r)$ is the mass of the halo inside radius $r$, 
as a ``proxy'' for mass.

Halos begin to form at very early epochs.  For example, at $z \sim 6$
we identify $>3,000$ halos with maximum circular velocity,
\Vmax, greater than $90~\kms$.  The numbers of halos
that we find at $z=3$, 1, and 0 are 14102, 14513, 10020, respectively.
We use a limit of $90~\kms$ on the circular velocity  
which is slightly lower than the completeness 
limit $\sim (110-120)\kms$ \citep{Colin98} of our halo catalog. 
This \Vmax value increases the number of halos
quite substantially (a factor of two as compared with the limit of $120~\kms$),
and, thus, reduces the statistical noise. 
We checked that our main results are only slightly affected 
by partial incompleteness of the sample.

Our halo identification algorithm, the Bound Density Maxima (BDM; see
KGKK), is described in detail elsewhere (Klypin \& Holtzman 1997).
The main idea of the BDM algorithm is to find positions of local
maxima in the density field smoothed at the scale of interest
($20 \kpch$). BDM applies physically motivated criteria to test whether
a group of DM particles is a gravitationally bound halo.  The major
virtue of the algorithm is that it is capable of finding both isolated
halos and halos orbiting within larger dense systems. Cluster-size
halos were also located by the BDM algorithm. The physical properties
of a sample of the twelve most massive groups and clusters\footnote{The
cluster number 8, in descending order in mass, was excluded from the
sample because it has a group close to it that produces too much
disturbance to the cluster.} are shown in Table 1. The total number of
clusters chosen for the sample is a compromise between taking a
relatively large number of clusters, so that we could talk about 
cluster average properties, and using clusters with a relatively high
number of halos. This cluster sample
is used to compute the average DM and halo velocity dispersion profiles as
well as the average DM and halo velocity anisotropy profiles.
\begin{deluxetable}{crrcr}
\tablecolumns{5}
\tablewidth{320pt}
\tablecaption{Physical Parameters of Clusters\tablenotemark{a}}
\tablehead{ \colhead{$M_{vir}$} & \colhead{$\sigma_{3D}$} & 
\colhead{$V_{max}$ } & \colhead{$R_{vir}$} & \colhead{$n_{halo}$} \\ 
\msunh & (km/sec) & (km/sec) & (\mpch) & $V_{\rm max}>90 \kms$ }
\startdata
$6.5 \times 10^{14}$ & 1645\phm{100} & 1402\phm{100} & 1.43 & 246\phm{100} \\ 
$2.4 \times 10^{14}$ & 1022\phm{100} & 910\phm{100} & 1.28 & 132\phm{100} \\
$1.9 \times 10^{14}$ & 992\phm{100} & 831\phm{100} & 1.17 & 98\phm{100} \\
$1.6 \times 10^{14}$ & 975\phm{100} & 789\phm{100} & 1.11 & 95\phm{100} \\
$1.4 \times 10^{14}$ & 887\phm{100} & 747\phm{100} & 1.05 & 58\phm{100} \\
$1.3 \times 10^{14}$ & 887\phm{100} & 730\phm{100} & 1.02 & 55\phm{100} \\
$1.1 \times 10^{14}$ & 831\phm{100} & 695\phm{100} & 0.98 & 45\phm{100} \\
$1.0 \times 10^{14}$ & 820\phm{100} & 680\phm{100} & 0.95 & 33\phm{100} \\
$9.9 \times 10^{13}$ & 789\phm{100} & 673\phm{100} & 0.94 & 74\phm{100} \\
$9.7 \times 10^{13}$ & 789\phm{100} & 668\phm{100} & 0.94 & 60\phm{100} \\
$9.3 \times 10^{13}$ & 753\phm{100} & 659\phm{100} & 0.92 & 67\phm{100} \\
$8.3 \times 10^{13}$ & 720\phm{100} & 635\phm{100} & 0.89 & 64\phm{100} \\
\tablenotetext{a}
 { Column description: (1) virial mass of the cluster; (2) 3D
velocity dispersion of dark matter particles; (3) maximum
circular velocity; (4) cluster radius; (5) number of
galaxy-size halos with $V_{max} > 90 \kms$.
 }
\enddata
\end{deluxetable}


\section{Results}

\subsection{The pairwise velocity bias}

The three-dimensional pairwise velocity dispersion  PVD is defined as
\begin{equation}
\sigma_{3D}^2(r) = \left<  \mathbf{v_{12}}^2 \right> - \left< \mathbf{v_{12}} \right>^2
\end{equation}
where $\mathbf{v_{12}}$ is the relative velocity vector of a pair of
objects separated by a distance $r$ and brackets indicate 
averaging over all pairs with the separation $r$. Figure 1 shows
the PVD for the dark matter, \sig3dm, at four epochs (top panel).  At 1
\mpch\ the radial PVD is about 1100 \kms\ at $z=0$. For the same
cosmological model Jenkins et al. (1998) find a radial PVD of $\sim 910
~\kms$. Jenkins et al. used slightly lower normalization for the model
($\sige = 0.9$) and used a bigger simulation box ($L_{box} = 141.3
\mpch$).  When the differences in
\sige\ are taken into account the Jenkins
et al. value increases to 1,120 \kms. Thus, both estimates roughly
agree. 
\begin{figure}[tb!]
\plotone{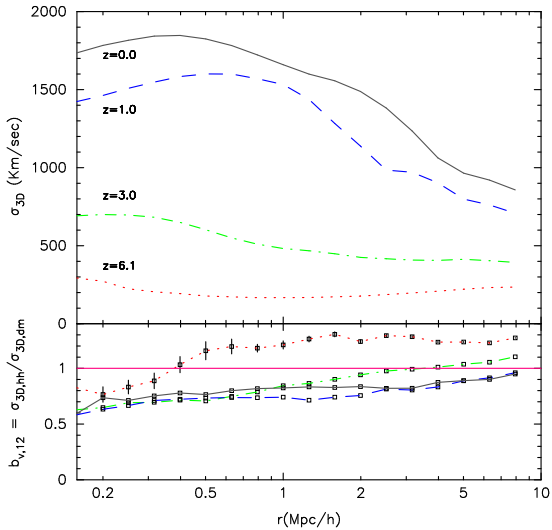}
\caption{\small {\it Top panel:} Three-dimensional pairwise
rms velocity  of the dark matter at four different epochs indicated in
the Figure.   {\it Bottom panel:} Pairwise velocity bias for galaxy-size
halos with circular velocities $V_{max} > 90 \kms$. Curves are labeled
in the same way as in the top panel. At very early
epochs and on large scales halos tend to move faster than the dark
matter. At later moments the pairwise velocity bias becomes smaller than 
unity.}
\end{figure}

The ratio of the halo and the dark matter PVDs, the pairwise velocity
bias $b_{v,12}$, is shown in the bottom panel of Figure 1.  All halos
with $V_{max} > 90~\kms$ were included in the computation.  At very
early epochs and on large scales halos tend to move faster than the
dark matter. At later moments the pairwise velocity bias becomes
smaller than unity.  It is interesting to compare the evolution of
$b_{v,12}$ with the changes in the spatial bias for the same
cosmological model \citep{Colin98}.  The spatial bias is defined as
the square root of the ratio of correlation functions $[\xi_{\rm
hh}(r)/\xi_{\rm dm}(r)]^{1/2}$. In general, the biases evolve in the
same way.  At high redshifts both biases are positive ($b>1$) and
decline as the redshift decreases. At low redshifts biases dive below
unity (antibias) and stop evolving. In spite of similarities, there
are some differences. The pairwise velocity bias becomes less than
unity at $z=3$ on scales below $3\Mpch$. At the same redshift the
spatial bias is still positive on all scales.

Col\'in \etal\/ (1999) and Kravtsov \& Klypin (1999) interpret the
evolution of the spatial bias as the result of several competing
effects. Statistical bias (higher peaks are more clustered) tends to
produce large positive bias and explain bias evolution at high
redshifts. At later epochs halos of a given mass or circular
velocity become less rare and start merging inside forming groups of
galaxies. Both effects lead to a decrease of bias. The merging becomes less
important as clusters with large velocity dispersions form at
$z<1$. This results in a very slow evolution of the halo correlation
function and bias. It is likely that the same processes define the
evolution of the pairwise velocity bias. The differences can be
explained by the known fact that the PVD is strongly dominated by few
largest objects (e.g., Zurek \etal\ 1994; Somerville, Davis \& Primack
1997): merging of halos inside forming groups at $z=3$ results in
fewer pairs with large relative velocities and in velocity antibias on
$\approx 1\Mpch$ scales. If this interpretation is correct, the
pairwise velocity bias mostly measures the spatial bias, not the
differences in velocities.

\subsection{The velocity anisotropy $\beta$}

A sample of 12 groups and clusters (see Table 1) was used to compute
various average cluster velocity statistics. In order to reduce the
noise in the profiles because of the small number of clusters in the
sample, we double the sample by using also the same clusters at slightly
different time $z=0.01$. For each cluster the halo distances to the
cluster center are divided by the corresponding cluster virial radius
(normalized distances). The halo velocities (averaged in spherical
bins) are divided by the corresponding cluster circular velocity at the
virial radius (normalized velocities). In Figure 2 we show radial
profiles, in normalized units, for halos and DM: the mean radial
velocity ($v_r$), the radial ($\sigma_r$) and the tangential
($\sigma_t$) velocity dispersions.  All halos are given equal
weight. We have accounted for the Hubble flow when we compute
$\sigma_r$ and $\sigma_t$ (so, proper, not peculiar velocities are
used); no correction for the mean radial velocity was made.  The trend
in both the velocity dispersion and the anisotropy velocity is slightly
affected if the mean radial velocity is subtracted at distances
$\simgreat 0.6$ and it is not affected at all at smaller distances.
\begin{figure}[tb!]
\plotone{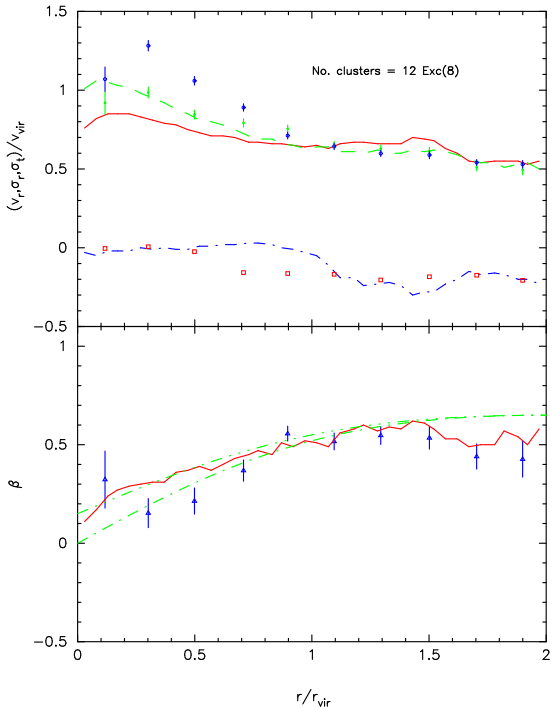}
\caption{\small Velocity profiles averaged over all clusters in
Table 1 excluding cluster \#8.  Curves are for the dark matter and
different symbols are for halos with circular velocity $V_{max} > 90
\kms$. Distances to the cluster centers are divided by the
corresponding cluster virial radius and halo velocities (averaged in
spherical bins) are divided by the corresponding cluster circular
velocity at the virial radius. Errorbars show 1-sigma errors of the
mean.  {\it Top panel:} Mean radial velocity (dot-dashed curve), the
radial velocity dispersion (full curve), and the tangential velocity
dispersion (dashed curve) of the dark matter.  Open squares, filled,
and open circles show the radial velocity, the radial velocity
dispersion, and the tangential velocity dispersion respectively for
halos.  {\it Bottom panel:} Velocity anisotropy for halos (open
triangles) and for the DM (solid line). The dot-dashed and
three-dot-dashed lines represent the fitting $\beta = 
{4r}\beta_m/(r^2+4) + \beta_0$ for two pairs of ($\beta_m$, $\beta_0$):  
(0.65,0) and (0.5,0.15), respectively.
}
\end{figure}

The velocity anisotropy function 
\begin{equation}
\beta = 1 - \sigma_t^2/2\sigma_r^2
\end{equation} 
\noindent is presented in the bottom panel of Figure 2 for halos
and for DM. For pure radial orbits $\beta = 1$, while an isotropic
velocity dispersion implies $\beta = 0$. The two lines added to the
panel show a fitting formula \citep{Carlberg97}.

\begin{equation}
 \beta = \beta_m \frac{4r}{r^2+4} + \beta_0
\end{equation}
\noindent for two pairs of parameters  
($\beta_m$, $\beta_0$): 
(0.65,0.) and (0.5,0.15).  
The first set of parameters gives a better approximation for halos. It
explicitly assumes that $\beta =0$ at the center. The second set of
parameters allows a small anisotropy at the center. It provides a
better fit for the dark matter. Note that while the halos have a
tendency for more isotropic velocities (with possible exception of the
center), the difference between halos and the dark matter is not
statistically significant.

 The variances of $\sigma_r$ and $\sigma_t$ are computed using
standard expressions for errors; for example, for $\sigma_r$
\begin{equation}
var(\sigma_r) = \frac{\mu_4 - \mu_2^2}{4n\mu_2},
\end{equation}
where $\mu_2 = \sum_{i} (v_{r,i} - \bar v_{r,i})^2$ and $\mu_4 =
\sum_{i} (v_{r,i} - \bar v_{r,i})^4$, and $n$ is the number of halos.
The statistical error is, thus, given by the square root of
$var(\sigma_r)$. The variance of $\beta$ is given by
\begin{equation}
[var(\beta)]^2 = \left( \frac{var(\sigma_t^2)}{2 \sigma_r^2} \right)^2 +
\left( \frac{var(\sigma_r^2)}{2 \sigma_r^4} \sigma_t^2 \right)^2.
\end{equation}

\subsection{The cluster velocity bias}

The three-dimensional velocity dispersions for both halos and DM are
shown in the top panel of Figure 3. The bottom panel shows the cluster
velocity bias, defined here as $b_{v} =
\sigma_{3D,halo}/\sigma_{3D,dm}$. It is surprising that halos in
clusters appear to have larger, by about 20\%, velocity dispersions
than the DM particles (positive bias). The trend is the same
regardless of what {\it type} of velocity dispersion (3D, tangential or
radial) we use in the velocity bias definition. There is almost
no bias in the very center of clusters. However, 
the $b_v$ value of the innermost bin increases if we exclude
the ``cD'' halos (defined as those halos which lie within
the inner $\sim 100 \kpch$ radius and have maximum circular
velocities greater than about 300 \kms). 
Their exclusion increases the positive velocity bias
to 1.22, a value which is comparable to that found in the adjacent bin.
\begin{figure}[tb!] 
\plotone{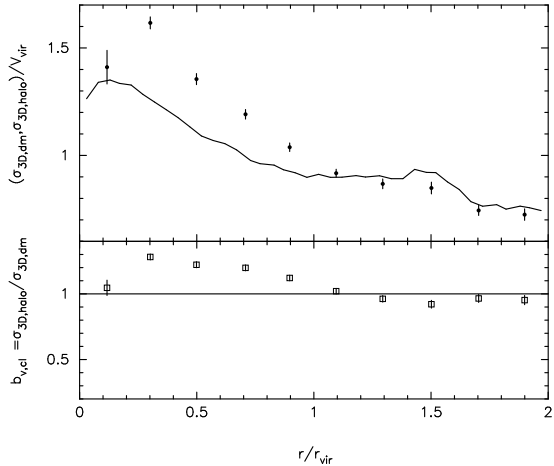}
\caption{\small {\it Top panel:}  3D velocity dispersion profiles
for halos (filled circles) and DM (solid line) in units of the mean
virial velocity. {\it Bottom panel:} Cluster velocity bias profile.
Errors correspond to 1-sigma errors of the mean.}
\end{figure}

The cluster positive velocity bias is robust to changes in the limit of
the circular velocity \Vmax. Only the innermost bin experiences
significant changes when this limit is increased. For example, when we
increase \Vmax~ from $90~\kms$ to $150~\kms$ (more massive halos are
chosen) the value of $b_v$ in the innermost bin reduces to 0.6.  This
favors a picture in which the central regions of clusters large
galaxy-size halos feel more the slowing effect of the dynamical
friction. All the other bins (within the virial radius) continue to
show small positive velocity biases.  The positive velocity bias is
also robust to changes in the number of clusters of the sample.  For
instance, one might suspect, that the most massive cluster weights so
much that it alters the statistics\footnote{In fact, the most massive
cluster of our simulation has had a recent major merger and halos may
still have large (``overheated'') velocities \citep[\eg][]{KW93}}. This
is true to some degree, but is {\it does not change} the ``sign'' of
the bias. For example, when we exclude this cluster and take $\Vmax
= 150~\kms$, all bins continue to show positive bias (within the virial
radius) except the innermost bin where $b_{v}= 0.5$. The results for
the innermost bin should be taken with caution because the effects of
overmerging may still be present in the central $100\kpch$ of the
clusters.

The difference in velocity dispersions of halos and dark matter
particles indicates that their velocity distribution functions (VDF)
are different.  We have examined both differential and cumulative VDFs
for the analyzed clusters and found that the halo VDFs are generally
skewed towards higher velocities as compared to the dark matter VDF, at
$r/r_{vir}\lesssim 0.8$.  The two VDFs are approximately the same for
larger radii. The observed differences in the velocity distribution may
be caused either by the differences in velocity fields of infalling
halos and dark matter (if, for example, halos are accreted
preferentially along filaments resulting in orbits of higher
ellipticity) or by effects of dynamical friction operating on halos,
but not on dark matter, in clusters. The dynamical friction may affect
the slowest halos more efficiently because the dynamical friction time
is proportional to the cube of the halo velocity. The slowest halos may
therefore merge more efficiently thereby skewing the velocity
distribution of the surviving halos towards higher velocities.

One could ask whether or not this positive cluster velocity bias
persists in the next set of twelve clusters or groups, in descending
order in mass (with virial masses below those clusters shown in Table
1).  Because this new sample of clusters have an average virial mass
lower than the average mass of clusters of Table 1, dynamical friction
is expected to operate more efficiently \citep[\eg][]{WR88, Diaferio98}.
The number of halos per cluster or group in this new sample is small,
we therefore use the whole group velocity dispersion. We find integral
$b_v$ values which are in general lower than one, and in some cases
there are groups that exhibit a strong velocity antibias (ratios close
to 0.6). This is contrary to what we find for the clusters of Table 1,
where the majority of clusters have an integral positive velocity bias.


\section{Discussion}

Literature on the velocity bias is very extensive and results are often
contradictory.  In this section we review some of the published results
and compare them with our results.  There are some reasons for the
chaotic state of the field. One of them is the confusion of two
different notions of the velocity bias -- the single-point $b_v$ and
the pairwise $b_{v,12}$ biases. The biases have different nature, and,
thus, give different results.  Another source of confusion is the way
how galaxies are identified or approximated in theoretical models. When
we combine this uncertainty with many physical processes, which we
believe can create and change velocity bias, the situation becomes
rather complicated.

{\bf Velocity profiles} seems to be the easiest part of the picture. In
this paper we present results, which are less noisy and are based on a
more homogeneous set of clusters than in most of previous publications.
Our results on the average cluster profiles for the dark matter ($v_r$
and $\sigma_r$) roughly agree with the results of \citet{CL96},
\citet{TBW97}, and \citet{Thomas98}. For example, \citet{TBW97} find a
DM velocity anisotropy $\beta_{dm} \simless 0.2$ at $r/r_{vir} \simless
0.1$ and $\beta_{dm} \simeq 0.5$ at $r/r_{vir} \sim 1$, which is close
to our results.  The structure of galaxy clusters in various
cosmologies is analyzed in detail by \citet{Thomas98}. From a total
sample of 208 clusters they choose a subsample which shows no
significant substructure. They find a more isotropic averaged beta
profile ($\beta_{dm} \sim 0.3$ at $r/r_{180} =1$) in their $\Lambda$CDM
model. The differences between our result and theirs can be accounted
for the fact that their clusters were selected not to have significant
substructure. More substructure in a cluster likely means a more
anisotropic cluster. The $\beta$ value at the cluster center (innermost
bins) is around 0.1, which is close to our results.

{\bf Pairwise velocity bias} is very sensitive to the number of pairs
found in rich clusters of galaxies. Removing few pairs may
substantially change the bias.  Thus, it mostly measures the spatial
bias (or antibias) and is less sensitive to real differences in
velocities.  The value of $b_{v,12}$ that we find at $z= 0$ is
typically higher than previous estimates reported in the literature,
computed for the $\Omega_0 = 1$ CDM model \citep{CC89, CCT90, GB94,
SDE95} Some of the results are difficult to compare because the
pairwise velocity bias is expected to evolve with time and vary from
model to model.

The first interesting result of this paper, that comes out from
the evaluation of $b_{v,12}$ at very high redshift, is that the halo
PVD {\it can be} greater than that of the DM. This positive velocity
bias had not been detected before (but see below) partly because of the
lack of simulations with very high resolution that could overcome the
overmerging problem. This result is surprising in part because halos
are expected to be born dynamically cool \footnote{Halos tend to form
near the peaks of the DM density distribution, 
\citep[\eg][]{Frenk88}.}. In fact, this is one of the reasons given in the
literature to explain the present-day pairwise velocity bias 
\citep[\eg][]{ESD94}. The other is the dynamical friction \citep[\eg][] {CCT90}. We
offer the following explanation to this positive velocity bias. Those
halos that are formed at very high redshift come from very high density
peaks. They are dynamically {\it cooler} than an average DM particle
from the region where they were born in, but {\it hotter} than most of the
matter.  The pairwise velocity bias $b_{v,12}$ rapidly becomes smaller
than one at non-linear scales.  As time goes on, the mergers inside
forming groups reduce the number of high velocity halos, while
velocities of DM particles increase.  As the result
the average halo random relative velocities are reduced below that of
the DM.

Using a semi-analytical method to track the formation of galaxies
Kauffmann et al. (1998a; 1998b) also find a pairwise velocity bias
greater than one at high redshifts.  They find that the galaxy PVD is
greater than that of the DM at $z > 1.1$ (their figure 11, $\tau$CDM
model). A $b_{v,12} > 1$ is expected at higher redshift in their
$\Lambda$CDM model as well.

{\bf Single-point velocity bias} appears to be the most difficult and
controversial quantity. It is important because it is a more direct
measure of the velocity differences. It still depends on the spatial
bias, but to much lesser degree as compared with the pairwise bias. An
interesting result was found when we evaluated the average cluster halo
velocity dispersion profile and compared it with that of the DM
particles: within the virial radius {\it halos move faster than the
dark matter}.

We believe that the explanation for this fact comes from a combination
of two known physical mechanisms: the dynamical friction and the
merging of halos. One may naively expect that the dynamical friction
should always slow down halos, which must result in halos moving slower
than the dark matter particles. This is not true. While on a short
time-scale the dynamical friction reduces velocity of a halo, the halo
may decrease or increase its velocity depending on the distribution of
mass in the cluster and on the trajectory of the halo. For example, if
a halo moves on a circular orbit inside a cluster with the
Navarro-Frenk-White profile, its velocity will first increase as it
spirals from the virial radius to $2.2R_s$, where $R_s\approx
(200-300)~{\rm kpc}$ is the characteristic radius of the core of the
cluster. The halo velocity will then decrease at smaller radii. When
the halo comes close to the center of the cluster it merges with the
central cD halo, which will have a tendency to increase the average
velocity of remaining halos. It appears that the Jeans equation
provides a better tool for understanding the velocity bias.
 
We will use the Jeans equation as a guide through the jungle of
contradictory results. It cannot be used more than a hint because it
assumes that a cluster is stationary and spherical, which is generally
not the case.  If a system is in a stationary state and is spherically
symmetric, the mass $M(<r)$ inside radius $r$ is related to the radial
velocity dispersion $\sigma_r$:
\begin{eqnarray}
M(<r) =& \frac{r \sigma_r^2}{G}A,\qquad\qquad\qquad \qquad\\
 A\equiv&  -\left( \frac{d \ln
\sigma_r^2}{d \ln r} + \frac{d \ln \rho}{d \ln r} + 2 \beta \right),
\end{eqnarray}
where $\rho$ is the (number) density profile, and $\beta$ is the
velocity anisotropy function.  The left-hand-side of this equation (the
total mass) is the same for both halos and the dark matter. Thus, if
the term $A$ is the same for the dark matter and halos, then there
should be no velocity bias: halos and the dark matter must have the
same $\sigma_r$. Numerical estimates of the term $A$ are inevitably
noisy because we have to differentiate noisy data. Nevertheless, we
find that the value of the term $A$ for halos is systematically smaller
than for the dark matter.  This gives a tendency for $\sigma_r$ to be
larger for halos.  In turn, this produces a positive velocity bias. The
main contribution comes typically from the the logarithmic slope of the
density: the halo density profile is shallower in the central part as
compared with that of the dark matter. The halo profile is shallower
likely because of merging in the central part of the cluster, which
gave rise to a central cD halo found in each of our clusters. We note
that while the Jeans equation shows the correct tendency for the bias,
it fails to reproduce correct magnitude of the effect: variations of
the term $A$ are smaller than the measured velocity bias.

One can also use the Jeans equation in a different way -- as an
estimator of mass.  We have computed $M(<r)$ for our average cluster
using both DM and halos.  At $\langle r/r_{vir}\rangle = 0.25$, where $b_v$ is close
to its maximum, the halo mass determination is larger than that of the
DM by a factor of 1.4.  This is due to the larger halo velocity
dispersion.  Because the term $A$ is actually higher for DM by about 10\%,
the overestimation is reduced from 1.56 to 1.4. As the distance to
the cluster center approaches the virial radius the mass overestimation
disappears.  At the virial radius both mass estimations agree,
essentially because $\beta$, $\sigma_r$, and the sum of the
logarithm derivatives are the same for both halos and DM, and are
within $(10-15)$\% of the true mass.

Using the Jeans equation for a spherically symmetric system and
assuming an isotropic velocity field, Carlberg (1994) showed that a
cool tracer population, $b_v < 1$, moving inside a cluster with a
power-law density profile (the density profile for the tracer is also
assumed to be a power-law), produced a mass segregation. That is, the
tracer population had a steeper density profile. We can invert this
reasoning and say that a more centrally concentrated halo distribution
produces a velocity antibias. We do not find this kind of mass
segregation in our halo cluster distribution. In fact, we see the
opposite -- halos are less concentrated than DM.  Dynamical friction
along with merging produces a lack of halos in the center of the
cluster. This very likely explains differences between our and
Carlberg's results for the velocity bias.

Carlberg \& Dubinski (1991) simulated a spherical region of 10 Mpc
radius and $64^3$ DM particles. They were unable to find galaxy-size
halos inside cluster at $z=0$ because of insufficient resolution:
softening length was 15~kpc instead of $\sim 2 \kpch$ needed for
survival of halos (KGKK). Their identification of
``galaxies'' with those DM particles which were inside high-density
groups found at high redshift, may have produced a spurious cluster
velocity antibias.  Using different galaxy tracers Carlberg (1994) also
found an integral cluster velocity bias lower than one. This result
could still be affected by numerical resolution ($\epsilon = 9.7
\kpch$). Evrard, Summers, \& Davis (1994) run a two-fluid simulation
in a small box, $L_{box} = 16~\mpc$, and stopped it at $z =1$.  Each DM
particle had a mass $9.7 \times 10^8
\msun$ and an effective resolution of 13~kpc (at $z =1$). The initial
conditions were constrained to assure that a poor cluster could form in
their simulation. Their ``globs'' (galaxy like objects) exhibit a lower
than one velocity bias. This velocity bias appeared not to depend on
epoch and mass. Their velocity antibias qualitatively agrees with our
results for groups and poor clusters.  At the same time, their value
for the pairwise velocity bias agrees with our results.

\citet{ME97} use an ensemble of two-fluid simulations to compute the
structure of clusters. Unfortunately, their simulations do not have
high mass resolution to allow the gas in their simulations to cool and
form ``galaxies'' (and then they could also allow for some
feedback). Instead, they use a high-density peak recipe to convert
groups of gas particles into galaxies particles. They find a one-point
``galaxy'' velocity bias that depends on cluster mass: the higher the
cluster mass is the higher the $b_{v}$ value.  We find a similar result
when we do the analysis of the velocity bias cluster by cluster
\footnote{On individual clusters we take only integral velocity
dispersions}. Their ensemble-averaged bias parameter is 0.84.  Their
recipe for galaxy formation produces a galaxy number density profile
which is steeper than that of the DM. This is likely the reason why
they find a $b_v$ value lower than one (Carlberg 1994, see above).

\citet{Frenk96} simulated a Coma-like cluster with a
P$^3$M + SPH code that includes the effects of radiative cooling. The
mass per gas particle is $2.4 \times 10^9\msun$ with a softening
parameter $\epsilon = 35~\kpc$ of the Plummer potential. Their galaxies
have two extreme representations: one as a pure gas clumps and
the other as lumps of the stellar component. They find a mass segregation
in both representations -- galaxies are more clustered than DM toward
the center of the cluster which is not seen in our halo distribution
\footnote{The reader might want to compare the Fig.  11 in Frenk et
al. (1996) with the Fig. 2 in Col\'in et al. (1999)}.  Once again,
according to Carlberg (1994) analysis, this would result in a one-point
velocity bias lower than one ($b_v \simeq 0.7$).  Because of a strong
cooling, their ``galaxies'' can acquire high density contrasts, which
helps galaxies to survive inside cluster.  At the same time, poor
force resolution (35~kpc) could have affected their results.

There are two studies where $b_v$ values greater than one are obtained.
Okamoto \& Habe (1999) simulate a spherical region of 30 Mpc radius
using a constrained random field method. They use a multi-mass initial
condition to reach high resolution. Their high-resolution simulated
region, where the cluster ends up, has a softening length $\epsilon =
5~\kpc$ and mass per particle $m \sim 10^9 \msun$. They find a cluster
velocity bias lower than one {\it only} in the innermost part of the
cluster where dynamical friction is expected to be more efficient. A
small positive bias ($b_v > 1$) is found in the range $0.3~\mpc < r
<0.6~\mpc$.  Based on the previous work by Kauffman et al. (1998a), Diaferio
et al. (1998) study properties of galaxy groups and clusters. They also find
that galaxies in clusters are ``hotter'' than the underlying dark matter 
field. They suggest that this effect is due to the infall 
velocities of blue galaxies. Infall could explain the positive velocity bias
of the outermost bin (within the virial radius) of our Figure 3, but
it definitely cannot account for the $b_v > 1$ value seen in the
inner bins (the mean radial velocity is close to zero for both DM and halos
in the three innermost bins).

There are several differences between our simulation and those
mentioned above. First, some of the papers cited above simulate only a
region which ends up as a cluster. So, they have structure for {\it
only} one cluster. The single-cluster one-point velocity bias could not
represent an {\it average} velocity bias, found using a sufficiently
large sample of clusters. For example, if our small positive velocity
bias is influenced by non-equilibrium cluster features, then when one
selects a cluster which is in {\it good} dynamical equilibrium (this
could be defined, for example, by the absence of substructure in the
cluster) and computes the one-point velocity bias, it could be biased
toward low values ($b_v < 1$) because dynamical friction have had
more time to operate.  Second, we simulate a relatively large random
volume that gives us many clusters in which effects such as tidal
torques, infall, and mergers are included naturally.  A cluster
simulated region or a random large region but without enough resolution
may not have a sufficiently large number of galaxy tracers and, thus,
introduce high statistical errors.  Our relatively large number of
halos in clusters reduces significantly the statistical errors in the
computation of $b_v$ and makes them suitable to the determination of,
for example, the radial dependence of the velocity bias.  Third, in
view of the Okamoto \& Habe (1999) and Ghigna et al. (1998) results,
and our own results, it seems that numerical resolution not only plays
an important role in determining the whole cluster velocity bias value
(both spatial and velocity bias intervene to affect its value) but it
is also important in determining the radial dependence of $b_v$ (almost
pure velocity bias).

What could account for the small positive velocity bias that we see in
our average cluster? We have examined both the differential and the
cumulative radial velocity distribution functions. We use the radial
velocity to highlight any contribution of infall velocities to the
velocity bias. The cumulative radial velocity distribution function is
shown in the Figure 4 for four different radial bins. In the top-right
panel (the innermost bin) we see a higher fraction of low-velocity halos
at small $v_r$ values.  This is due to central cD halos, which move very
slowly relative to clusters themselves.  At large $v_r$ values we
observe the contrary -- a higher slope, which means that there are many
fast moving halos. If we do not include the cD halos, the velocity bias
becomes larger then unity even in the central radial bin.  However, as
we noticed earlier in section 3.3, a velocity antibias can appear in
the central bin, if the value of \Vmax~ is increased. It is clear that
the deficiency of low and moderate $v_r$ halos produces the positive
velocity bias measured at $r=(0.2-0.8) r_{vir}$ (see the top-left and
the bottom-right panels). We have used the Kolmogorov-Smirnov test in
order to evaluate whether or not the halo and the DM velocity
distribution functions are statistically different.  We find that the
probability that these functions were drawn from the same distribution
is smaller than 0.01 in all radial bins that are within the virial
radius.  As mentioned in
\S~3.3, the dynamical friction may have affected the slow moving halos
more significantly because the dynamical friction time-scale is
proportional to the cube of the halo velocity.  It is thus expected
that low-velocity halos merge sooner than their high-speed
counterparts, thereby skewing the VDF toward high-velocity
halos\footnote{It should be also kept in mind that as halos move to
orbits of smaller radii they could acquire higher velocities because
the DM velocity dispersion increases toward the cluster center}.
Infall could also be an important source of positive velocity bias for
the outermost bins.
\begin{figure}[tb!] 
\plotone{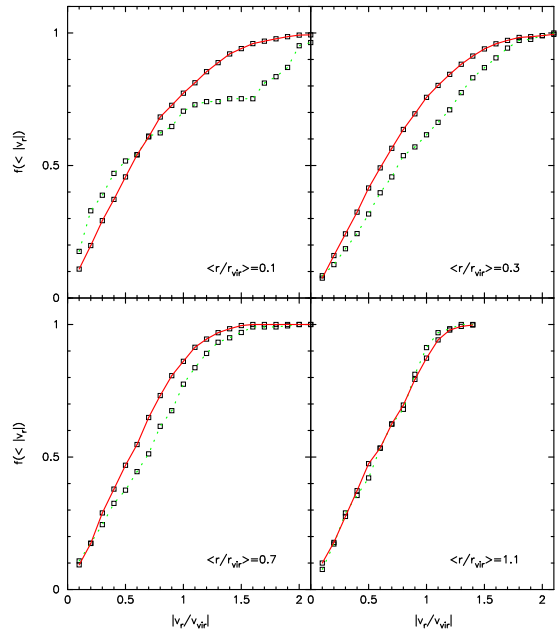} 
\caption{\small Cumulative radial velocity distribution functions
in four radial bins for halos (dotted line) and DM (solid line).}
\end{figure}

\section{Summary}

1. We have found that galaxy-size halos have a time- and scale-
dependent pairwise velocity bias. At high redshifts ($z \sim 5$) this
bias is larger than unity ($\approx 1.2$). It declines with time and
becomes $\approx 0.6-0.8$ at $z=0$.  The evolution of the pairwise
velocity bias follows and probably is defined by the spatial bias of
the dark matter halos.  These results are in qualitative agreement with
those by Kauffmann et al. (1998b).

2. We have evaluated the velocity anisotropy function $\beta(r)$ for
both halos and DM particles. For both halos and DM $\beta$ is a
function that increases with radius and reaches a value of $\simeq 0.5$
at the virial radius. The difference between this value and that found
by \citet{Thomas98} likely can be explained by the fact that
\citet{Thomas98} selected a sample of clusters which had little
substructure. Our simulations indicate that the halo velocity anisotropy
closely follows (but lies slightly below) that of the underlying dark
matter.

3. Halos in our clusters
move faster than DM particles: $b_v =(1.2-1.3)$ for
$r=(0.2-0.8)r_{vir}$. This result disagrees with many previous
estimates of the cluster velocity bias. This difference appears to be
due to differences in numerical resolution. 
More work needs to be done to settle the issue. Nevertheless, it is
encouraging that Diaferio et al. (1998) and Okamoto \& Habe (1999)
found results similar to ours. 

4. Usual argument that dynamical friction slows down galaxies and,
thus, must produce velocity antibias is not correct. Galaxy tracers in
clusters move through an environment which has a steep density
gradient. A sinking halo may either {\it increase} or decrease its
velocity depending on the distribution of cluster mass and on the
trajectory of the halo.
A combination of the dynamical friction and merging appears as the most
compelling hypothesis which could account for our small positive
velocity bias.

We acknowledge the support of the grants NAG-5-3842 and
NST-9802787. Computer simulations were performed at NCSA.
P.C. was partially supported by DGAPA/UNAM through project
IN-109896.



\begin{thebibliography}{DUM}

\bibitem[Benson et al. (1999)]{Benson99}
Benson, A.J., Cole, S., Frenk, C.S., Baugh, C.M., \& Lacey, C.D.
1999, MNRAS submitted (astro-ph/9903343)
\bibitem[Carlberg (1994)]{Carlberg94}
Carlberg, R.G. 1994, ApJ, 433, 468
\bibitem[Carlberg \& Couchman (1989)]{CC89}
Carlberg, R.G., \& Couchman, H.M.P. 1989, ApJ, 340, 47
\bibitem[Carlberg, Couchman \& Thomas (1990)]{CCT90}
Carlberg, R.G., Couchman, H.M.P., \& Thomas, P.A. 1990,
ApJ, 352, L29
\bibitem[Carlberg \& Dubinski (1991)]{CD91}
Carlberg, R.G., \& Dubinski, J. 1991, ApJ, 369, 13
\bibitem[Carlberg et al. (1996)]{Carlberg96}
Carlberg, R.G., Yee, H.K.C., Ellingson, E.,
Abraham, R., Gravel, P., Morris, S., Pritchet, C.J.
1996, ApJ, 462, 32
\bibitem[Carlberg et al. (1997)]{Carlberg97}
Carlberg, R.G. et al. 1997, ApJ, 485, L13
\bibitem[Cen \& Ostriker (1992)]{CO92}
Cen, R., \& Ostriker, J.P. 1992, ApJ, 399, L113
\bibitem[Chaboyer (1998)]{Chaboyer98}
Chaboyer, B. 1998, Phys. Reports, in press (astro-ph/9808200)
\bibitem[Cole \& Lacey (1996)]{CL96}
Cole, S., \& C. Lacey. 1996, MNRAS, 281, 716
\bibitem[Col\'in, Carlberg, \& Couchman (1997)]{CCC97}
Col\'in, P., Carlberg, R.G., \& Couchman, H.M.P. 1997,
ApJ, 490, 1
\bibitem[Col\'in et al. (1999)]{Colin98}
Col\'in, P., Klypin, A.A., Kravtsov, A.V., \& A.M. Khokhlov. 1999,
ApJ in press (astro-ph/9809202)
\bibitem[Couchman \& Carlberg (1992)]{CC92}
Couchman, H.M.P., \& Carlberg, R.G. 1992, ApJ, 389, 453
\bibitem[Davis et al. (1985)]{DEFW85}
Davis, M., Efstathiou, G., Frenk, C.S., \& White, S.D.M. 1985,
ApJ, 292, 371
\bibitem[Davis \& Peebles (1983)]{DP83}
Davis, M., \& Peebles, P.J.E. 1983, ApJ, 267, 465
\bibitem[Diaferio et al. (1998)]{Diaferio98}
Diaferio, A., Kauffmann, G., Colberg, J.M., \& White, S.D.M. 1998,
MNRAS submitted (astro-ph/9812009)
\bibitem[Eke et al. (1998)]{Eke98}
Eke, V.R., Cole, S., Frenk, C.S., \& Henry, P.J. 1998, MNRAS, 298, 1145
\bibitem[Evrard (1997)]{Evrard97}
Evrard, A.E. 1997, MNRAS, 292, 289
\bibitem[Evrard, Summers \& Davis (1994)]{ESD94}
Evrard, A.E., Summers, F.J., \& Davis, M. 1994, ApJ, 422, 11
\bibitem[Frenk et al. (1988)]{Frenk88}
Frenk, C.S., White, S.D.M., Davis, M., \& Efstathiou, G. 1988, ApJ, 327, 507
\bibitem[Frenk et al. (1996)]{Frenk96}
Frenk, C.S., Evrard, A.E., White, S.D.M., \& Summers, F.J. 1996, ApJ,
472, 460
\bibitem[Gelb \& Bertschinger (1994)]{GB94}
Gelb, J.M., \& Bertschinger, E. 1994, 436, 491
\bibitem[Ghigna et al. (1998)]{ghigna98}
Ghigna, S., Moore, B., Governato, F., Lake, G., Quinn, T., Stadel, J.
1998, MNRAS, 300, 146
\bibitem[Jenkins et al. (1998)]{Jenkins98}
Jenkins et al. (The Virgo Consortium) 1998, ApJ, 499, 20
\bibitem[Jing \& B\"{o}rner (1998)]{JB98}
Jing, Y.P.,\& B\"{o}rner, G. 1998, ApJ, 503, 502
\bibitem[Jing, Mo \& B\"{o}rner (1998)]{JMB98}
Jing, Y.P., Mo, H.J., \& B\"{o}rner, G. 1998, ApJ, 494, 1
\bibitem[Katz \& White (1993)]{KW93}
Katz, N., \& White, S.D.M. 1993, ApJ, 412, 455
\bibitem[Kauffmann et al. (1998a)]{Kauffmann98a}
Kauffmann, G., Colberg, J.M., Diaferio, A., \& White, S.D.M.
1998a, MNRAS, 303, 188
\bibitem[Kauffmann et al. (1998b)]{Kauffmann98b}
Kauffmann, G., Colberg, J.M., Diaferio, A., \& White, S.D.M.
1998b, MNRAS submitted (astro-ph/9809168)
\bibitem[Klypin et al. (1998)]{KGKK98}
Klypin, A., Gotl\"{o}ber, S.,  Kravtsov, A., Khokhlov, A.. 1998, 
ApJ, 516, 530 (KGKK)
\bibitem[Klypin \& Holtzman (1997)]{KH97}
Klypin, A., \& Holtzman, J. 1997, preprint (astro-ph/9712217)
\bibitem[Klypin \etal (1993)]{KHPR}
Klypin, A., Holtzman, J., Primack, J., \& Regos, E. 1993, ApJ, 416, 1.
\bibitem[Kravtsov \& Klypin (1999)]{KK98}
Kravtsov, A.V., \& Klypin, A. 1999, ApJ in press (astro-ph/9812311)
\bibitem[Kravtsov \& Klypin (1997)]{KKK97}
Kravtsov, A.V., Klypin, A., \& Khokhlov, A.M. 1997, ApJS 111, 73
\bibitem[Landy, Szalay, \& Broadhurst (1998)]{LSB98}
Landy, S.D., Szalay, A.S., \& Broadhurst, T.J. 1998,
ApJ, L133
\bibitem[Metzler \& Evrard (1997)]{ME97}
Metzler, C., \& Evrard, A.E. 1997, ApJ, submitted (astro-ph/9710324)
\bibitem[Okamoto \& Habe (1999)]{OH99}
Okamoto, T., \& Habe, A. 1999, ApJ, 516, 591
\bibitem[Peebles (1980)]{Peebles80}
Peebles, P.J.E. 1980, The Large-Scale Structure of the
Universe (Princeton: Princeton Univ. Press)
\bibitem[Roos \& Harun-or-Rashid (1998)]{RH99}
Roos, M., \& Harun-or-Rashid, S.M. 1998, A\&A, 329, L17
\bibitem[Somerville, Davis \& Primack (1997)]{Somerville}
Somerville, R., Davis, M., \& Primack, J. 1997, ApJ, 479, 616
\bibitem[Summers, Davis, \& Evrard (1995)]{SDE95}
Summers, F.J., Davis, M., \& Evrard, A.E. 1995, ApJ, 454, 1
\bibitem[Thomas et al. (1998)]{Thomas98}
Thomas et al. (The Virgo Consortium) 1998, MNRAS, 296, 1061
\bibitem[Tormen, Bouchet, \& White (1997)]{TBW97}
Tormen, G., Bouchet, F.R., \& White, S.D.M. 1997, MNRAS,
286, 865
\bibitem[West \& Richstone (1988)]{WR88}
West, M.J., \& Richstone, D.O. 1988, ApJ, 335, 532
\bibitem[Zurek \etal (1994)]{Zurek}
Zurek, W., Quinn, P.J., Salmon, T.K., \& Warren, M.S. 1994, ApJ, 431, 559
\end{thebibliography}
\end{document}